\documentclass[oneside]{amsart}

\usepackage{cancel}

\usepackage[utf8]{inputenc}
\usepackage[T1]{fontenc}

\usepackage[a4paper]{geometry}
\usepackage{parskip}

\usepackage{enumitem}

\usepackage{amsmath,amssymb,amsfonts}
\usepackage{mathtools}
\usepackage{graphicx}
\usepackage[font=small,labelfont=bf]{caption}
\usepackage{subfigure}
\usepackage{tikz}
\usetikzlibrary{calc,patterns}
\usetikzlibrary{math}
\usetikzlibrary{intersections}
\usetikzlibrary{scopes}
\usetikzlibrary{shapes}
\usetikzlibrary{arrows,automata}
\usepackage{pgfplots}
\pgfplotsset{                compat=1.11,
}

\usepackage{cancel}
\usepackage{color}

\usepackage{hyperref}
\hypersetup{
colorlinks=true,
linktocpage=true,
pdfstartview=FitH,
breaklinks=true,
pdfpagemode=UseNone,
pageanchor=true,
pdfpagemode=UseOutlines,
plainpages=false,
bookmarksnumbered,
bookmarksopen=false,
bookmarksopenlevel=1,
hypertexnames=true,
pdfhighlight=/O,
pdftitle={},
pdfauthor={},
pdfsubject={},
pdfkeywords={},
pdfcreator={pdfLaTeX},
pdfproducer={LaTeX with hyperref}
}

\DeclarePairedDelimiter{\braces}{\{}{\}}
\DeclarePairedDelimiter{\bracks}{[}{]}
\DeclarePairedDelimiter{\parens}{(}{)}
\DeclarePairedDelimiter{\abs}{\lvert}{\rvert}
\DeclarePairedDelimiter{\angbra}{\langle}{\rangle}
\DeclarePairedDelimiter{\supnorm}{\lVert}{\rVert_\infty}

\DeclareMathOperator{\eff}{eff}
\DeclareMathOperator{\dmu}{d\mathrm{\mu}}
\DeclareMathOperator{\argph}{\cdot}
\newcommand{\de}[1]{\operatorname{d}\!{#1}}
\newcommand{\allgraphs}[2][]{\mathcal{G}_{#2}^{#1}}
\newcommand{\congraphs}[1]{G_{#1}}

\newcommand{\edge}[1]{\braces{#1}}

\newcommand{\epsi}{\varepsilon}

\theoremstyle{plain}
\newtheorem{theorem}{Theorem}

\newtheorem*{corollary*}{Corollary}
\newtheorem{lemma}[theorem]{Lemma}

\theoremstyle{definition}
\newtheorem{definition}[theorem]{Definition}
\newtheorem*{definition*}{Definition}

\newtheorem*{hypothesis*}{Hypothesis}

\theoremstyle{remark}
\newtheorem{remark}[theorem]{Remark}
\newtheorem*{remark*}{Remark}
\newtheorem*{notation*}{Notational remark}

\newtheorem*{case*}{Case}

\numberwithin{theorem}{section}
\numberwithin{example}{section}

\title
[Lonely planets and light belts]
{Lonely planets and lightweight asteroids: a statistical
mechanics model for the planetary problem.}

\author
[G.~Pinzari]
{Gabriella Pinzari$^{\ddag}$}
\address{$^{\ddag}$ Dipartimento di Matematica ``Tullio Levi--Civita'', Università degli Studi di Padova, Via Trieste, 63, 35131 Padova, Italy}

\author
[B.~Scoppola]
{Benedetto Scoppola$^{\S}$}
\address{$^{\S}$ Dipartimento di Matematica, Università degli Studi di Roma ``Tor Vergata'', Via della Ricerca Scientifica, 1, 00133 Roma, Italy.}

\author
[A.~Troiani]
{Alessio Troiani$^{\star}$}
\address{$^{\star}$ Dipartimento di Matematica ``Tullio Levi--Civita'', Università degli Studi di Padova, Via Trieste, 63, 35131 Padova, Italy}
\email{alessio.troiani@math.unipd.it}

\begin{document}

\begin{abstract}
In this paper we propose a notion of stability, that we call $\epsi -N$-stability, for systems of particles interacting via Newton's gravitational potential, and orbiting a much bigger object. For these systems the usual thermodynamical stability condition, ensuring the possibility to perform the thermodynamical limit, fails, but one can use as relevant parameter the maximum number of particles $N$ that guarantees the $\epsi -N$-stability. With some judicious but not particularly optimized estimates, borrowed from the classical theory of equilibrium statistical mechanics, we show that our model has a good fit with the data observed in the Solar System, and it gives a reasonable interpretation of some of its global properties.\\

\end{abstract}

\maketitle

\section{Introduction}
Maybe  the secret of the huge success of  Kolmogorov--Arnold--Moser ({\sc kam}) theory  relies in the
spectacular application, found out by  Vladimir Igorevich Arnold,  to the planetary problem.
Indeed, one decade after Kolmogorov's announcement, at the 1954's International Congress of Mathematician, of the ``theorem of the conservation of the invariant torus'' \cite{Kolmogorov54}, the brilliant student of Kolmogorov -- aged 27 -- formulated a version of Kolmogorov's theorem (which he called the ``Fundamental Theorem'') suited to  the planetary problem \cite{arnold63}. He then used it  to prove the ``metric stability'' of the simplest, albeit non--trivial,  planetary system: two planets and a sun constrained on a plane. Strong degeneracies prevented a straightforward application of the Fundamental Theorem to the most general planetary system, which indeed was obtained in the subsequent 50 years, after those degeneracies were completely understood \cite{laskarR95, fejoz04, pinzari-th09, chierchiaPi11b, pinzari18}; see \cite{chierchiaPi14} for a review. The success of {\sc kam} theory in classical mechanics boosted  other investigations, like instability or finite time stability \cite{arnold64, nehorosev77}.

Despite the quality and the quantity of results of this kind, some intriguing questions remain open. In this paper we are particularly interested in the problems that have a kind of ``global'' structure in planetary systems, like, e.g., the mass distribution and/or the stability of the belts of many lightweight objects. In our Solar system this could be relevant to understand the global features of the various asteroids belts, the rings around the planets, or the space debris .

From this respect the basic idea of statistical mechanics, that is the possibility to substitute the exact knowledge of the dynamics of an $N$-particles system with a probability distribution on its dynamical status in a fixed instant, seems to be promising. Losing the detailed knowledge of the trajectory of the system in the phase space one gains the possibility to describe global quantities like e.g. pressure, temperature or density. With this attitude, the study of the analyticity of such global functions may give a microscopical justification of many very interesting phenomena, e.g. the phase transitions.

To fulfill this program, however, one has to assume various technical conditions, and one of the first constraints, understood from the very beginning of the discipline, is the so called stability condition on the interacting potential. Namely, one has to impose to the potential $V(x,y)$ of the interaction between particles the following condition: it has to exist a positive constant $B$ such that
\begin{equation}\label{boundedness}
\sum_{1\le i< j\le N}V(x_i,x_j)\ge  -BN
\end{equation}
for all the possible choices of the positions $x_1,...,x_N$ of the $N$ particles (see for instance \cite{Ruelle} and \cite{Galla}).
Such stability condition, obviously, does not hold for the Newtonian potential. In the quantum case it has been possible to show that a non stable potential, like the Coulomb one, leads to a stable behavior of the matter (see \cite{LL72}). Many other results (see e.g., \cite{Thirring71, LiebYau87, LiebYau}), have been proved exploiting explicitly the features of the Fermi and Bose statistics. In the study of classical systems with Newtonian interaction a possible way to perform the thermodynamic limit is to define interactions depending on the number of particles, the so-called mean field systems, or, almost equivalently, to rescale suitably the energy with the number of particle in a fixed volume, the so-called Vlasov limit \cite{MesSpo, Kies89, Kies09}.
This approach proves useful to model, for instance, plasma physics and astronomical globular systems.
In a different context, namely that of growing planets, some analytical results have been obtained recently by \cite{Tremaine} in the case of the planetary systems in which the instability is part of the desired result. In \cite{Tremaine2} and \cite{Roupas} the statistical mechanics of dense stellar clusters have been studied, outlining its similarity with liquid crystals.
Moreover, numerical simulations of gravitating systems represent a very active research field. 
Finally, other recent attempts to perform, numerically, a statistical analysis of the future planet orbits in the solar system include, e.g., \cite{Mogavero}. 

In this paper we propose a different approach, not yet previously investigated, 
at least to our knowledge. 
We study a model of interacting particles in a planetary system.
Our estimates are mathematically rigorous, although some of the assumptions underlying the
definition of the model are
motivated by heuristic considerations based on the observed features of the solar system.
Some of these features are suitably simplified.
The first simplification is the fact that our planetary system is assumed to be planar, and the central star is fixed. It will be clear that this assumption simplifies the details but it does not modify the structure of the bounds we will present. Secondly, we  deliberately decide to lose details in the description of the interaction between the center, called hereafter the star, and each light particle, call hereafter  asteroid. In particular, we describe each orbit as a probability distribution around a fixed circular orbit. Such probability distribution does not fix the energy of each asteroid. In other words, we try a description of the system in terms close to the familiar idea of canonical distribution. From a physical point of view, this assumption can be justified thinking for instance to the main belt of asteroids: the 2-body elliptical trajectory is actually an approximation due to the fact that each asteroid is perturbed by the planets. Hence the energy of the single asteroid is not conserved. We substitute the computation of the actual trajectory, perturbed by the planets, with the probability distribution mentioned above. 
The distribution described so far play the role of the reference, or free, measure, in the sense that each asteroid has its independent (i.e. factorized) free measure. Then we introduce an interaction in the probability distribution, adding a gravitational potential among asteroids. As far as the short distance configurations are concerned, a regularization in terms of hard core interaction among asteroids is introduced: if $x_i$ represent the vector position of the $i$-th asteroid and $a_i$ is its radius, then $|x_i-x_j|\ge a_i+a_j$, or, in other way, the potential $V_{ij}$ is infinite if $|x_i-x_j|\le a_i+a_j$. Note that in this context the planets are much bigger than the interacting asteroids, but are very far. The interacting asteroids, on the other side, are light but in principle they can have very small mutual distances, exactly of the order of the sum of their radii, and these colliding configurations will give a huge contribution to the interacting probability measure. We are not assuming in this model neither agglomerations nor disintegrations due to these collisions, we simply compute their static contributions to the interacting probability measure.
Even with this strong simplifications, however, it is hopeless to perform the thermodynamical limit. The hard core interaction prevents the possibility of configurations having an infinite probabilistic weight, but the features of the Newtonian interaction, and in particular its very slow decay, do not allow an estimate of the form (\ref{boundedness}). Nevertheless, assuming that the number of asteroids $N$ is a large but finite parameter, and discussing its value in terms of the masses of the asteroids, we find results that are quite interesting in terms of the description of the real Solar System.

In order to quantify the effect of the gravitational interaction on the trajectories of the asteroids we define a notion of stability in the following way: each asteroid, with respect to its independent reference measure, has its own variance of the distance from the star. Call $\sigma^2_0$ such variance. If the number $N$ of asteroids (and their masses distribution) is appropriately chosen it is possible, uniformly in the choice of the asteroid, to give for the interacting measure an estimate of the variance $\sigma^2$ of the form
\begin{equation}
\sigma^2=\sigma^2_0(1+\epsi)
\end{equation}
we then say that the system is $\epsi -N$-stable. If $\epsi$ is sufficiently small we can argue that the interaction among asteroids implies small modifications  of the asteroid's orbit. 

To give an initial idea of the smallness required on $\epsi$ in order to have an astronomical interpretation of this notion of stability, we are assuming that the energy of each asteroid confines it in the vicinity of the minimum of its effective potential. This means that the eccentricity is small, together with the effect of the interaction with the distant planets, and hence the reference (free) probability of each asteroid has a standard deviation around its average radius very small with respect to the radius itself. If the effect of the interaction among asteroids is such that this standard deviation stays small, i.e. if $\epsi$ is of order 1 or less, we can assume that the probability of a large deviation of an asteroid from its average radius is very small. 
A quite important point is the relation between this large deviation probability and the astronomical stability of the orbit.
The inverse of the probability of a large deviation multiplied by the time scale of the variations of the distance between the asteroid and the star, that is at least the period of the orbit of the asteroid itself, gives us an idea of the time scale in which we have to assume that such large deviation is not realized, and hence the asteroid will remain close to its present orbit. Such large deviation probabilities estimates could be done with rough but robust tools, like Chebyshev inequality, or with more sophisticated techniques. This delicate point will be discussed in the next section.

We study in details three different setup:
\vglue.3truecm
1) Similar asteroids

Our first setup is somehow theoretical: the asteroids are very light and their masses are comparable. The average radius of the orbit is similar for each asteroid; we will compute the stability of the system in the worst case, i.e. for equal average radii. Fixing the radius of the asteroids we find an estimate, depending on $\epsi$, of the maximum value of the number of asteroids $N$ such that the system is $\epsi -N$-stable. This first computation is important in order to understand that in this context it is not possible to perform a standard thermodynamical limit. However, if the total mass of asteroids depends on their number $N$, and it goes suitably to zero when $N$ increases, then it is possible to prove the $\epsi -N$-stability of the system. In this context it is easier to outline the basic problem that one has to face: we want to fix the parameter of the system in such a way that the contributions of the collisions, in which the Hamiltonian is negative and has a large absolute value, are not too relevant for the canonical probability distribution.

\vglue.3truecm
2) Asteroids with a given mass distribution

In this setup the asteroids have always a similar average distance from the star, but they have a well defined distribution of the masses. We show that a distribution of the form
\begin{equation}
N(>r)=\frac{c}{r^\nu}
\end{equation}
where $N(>a)$ is the number of asteroids with radius greater than $a$, $\nu>1$ and $c$ is a suitable positive constant, guarantees that $N$ can be chosen quite large, and yet the system remains $\epsi -N$-stable. Remarkably, our assumption about the radii distribution of the asteroids seems to be quite close to the observed one. In particular, the estimates obtained form observed data give a value of $\nu$ between $1.3$ and $3$ (see  \cite{Ryan} and references therein). Again, the basic problem is to control the contributions of the collisions. 

\vglue.3truecm
3) Planets with well separated orbits.

In the last part of this paper we try to apply the same techniques developed for asteroids to a system of planets, i.e. of object small with respect to the star but larger than asteroids, having orbits with very different average radii: we show that assuming for the average radius of the orbit of the $i$-th planet the following Titius-Bode law:
\begin{equation}
R_i=b+c a^i
\end{equation}
with $R_i$ the orbit's average radius of the $i$-th planet, $b$ and $c$ fixed length ($0.4$ and $0.3$ $U.A.$ respectively for the Solar System) and $a>1$ a fixed number ($a=2$ for the Solar System), and assuming $N$ small enough, the system is 
$\epsi -N$-stable. In this context we briefly discuss the $1$-stability of the Galilean Jupiter's satellites. Also in this case the main problem is to control the contributions due to the collisions. Here, however, we have to exploit the fact that in order to have a collision the planets have to deviate substantially from their reference distribution, see further comments below, in the beginning of section 5.

\vglue.3truecm
In order to obtain these results, we have to interpret the classical meaning of the thermodynamic constants in Gibbs distribution in a different way. 
The main problem is the interpretation of the temperature in this context. As it will be clear in the next section, it is physically meaningless to define a common temperature for objects with different masses. Since the exact Keplerian orbit and also its correction due to external objects are computed in terms of gravitational interaction, the shape of the orbit and, consequently, the form of the probability distribution that we want to define may not depend on the mass of the orbiting object. On the other side the contribution to the energy of such object depends linearly on the mass. Hence the contribution of an object of mass $m$ to the probability distribution has to be rescaled by a factor $1/m$. This means that each object has its own ``temperature'', proportional to $1/m$.
After this rescaling, it is natural to assign the role of the temperature to a number $\gamma$ that is related to the free measure deviation of the radii of the asteroids. Small temperature, corresponding in our model to large $\gamma$, means that the asteroid has a free distribution concentrated in the vicinity of its reference circular orbit. Hence we are assuming that in the low temperature regime the eccentricity of the orbits and the interactions with the heavy far planets are small, and the asteroids, as far as their free measure distribution is concerned, have an average energy very close to the minimum of their effective potential.

Moreover our particles are obviously distinguishable, and hence the combinatorial Gibbs factors $\frac{1}{N!}$ are absent in our treatment. In order to estimate the deviations of the radii of the asteroids (and, eventually, planets) in presence of the interaction among them we had to use a procedure quite standard in statistical mechanics, usually known as Peierls argument, see section 2 below, and then judicious combinatorial estimates, similar to the ones introduced in cluster expansion. Such estimates, due to the absence of the Gibbs factor, have some nonstandard features.

The results we obtain, despite the simplicity of the estimate we present, may have some interest. From a  quantitative point of view the estimates of the mass and of the number of the asteroids and of the planets are quite different from the ones observed in the Solar System, but the orders of magnitude are not too distant. In the simpler case of Galilean satellites our notion of stability is guaranteed for masses of the satellites close to the actual ones.
Moreover the model explains why in order to have stability the number of very light asteroids may be relatively high, while the planets have to be quite far apart and their number has to be very small, of the order of $N\le10$.

The model, then, seems to have a reasonable fit with the observed data.

The work is organized as follows: in section 2 we present our model, we define more precisely the notion of ``thermodynamical''  stability for planetary systems and we define the relation among this notion of stability and the astronomical one; in section 3 we discuss an application of the model to a belt of asteroids having a very narrow distribution of masses; in section 4 we generalize the same results to a more realistic asteroid belt; section 5 is devoted to the application of our model to a planetary system in which the radii of the planets satisfy a kind of Titius-Bode law. Finally in section 6 we discuss some brief final remarks.

\section{The model}
\label{sec:definition}

\subsection{Planetary system}

Consider a system of $N$ bodies with mass $m_i$, constrained on a plane, with pairwise gravitational interaction and interacting gravitationally with a much larger body, the star, of mass $M$ centered at the origin of a reference frame in the plane.  The system is described by the Hamiltonian
\begin{align}\label{eq:original_hamiltonian}
        H(\vec{p}, \vec{q}) = \sum_{i = 1}^{N} \frac{|p_{i}|^{2}}{2 m_i} -
                \sum_{i=1}^{N} \frac{k M m_{i}}{\abs{q_i}} -
                \sum_{1\le i<j\le N} \frac{k m_{i} m_{j}}{\abs{q_{i}-q_j}}
\end{align}
where $q_i$ are 2-dimensional euclidean coordinates, $p_i$ the corresponding moments and $k$ the gravitational constant.
We remark that in our model  the mass $M$ does not move. This appears in (\ref{eq:original_hamiltonian}) from having neglected centrifugal terms coming from taking  the reference frame centered at $M$; compare, e.g. \cite{fejoz04}, for the general expression of the $N$--body Hamiltonian in the star--centred frame.

Calling
\begin{align}
        H_{0}(\vec{p}, \vec{q}) = \sum_{i = 1}^{N} \frac{|p_{i}|^{2}}{2 m_i} -
                \sum_{i=1}^{N} \frac{k M m_{i}}{\abs{q_i}}
\end{align}
the Hamiltonian describing $N$ uncoupled central interactions with the star, the original Hamiltonian can be seen as the sum of $H_{0}$ and a perturbing term.

Rewriting each term $h_0$ of the sum appearing in $H_{0}$ using polar coordinates (with $\rho$ the distance from the star and $\theta$ the true anomaly)
and setting $p_{\theta} = J$
for the conservation of the angular momentum in the central system, we can write $h_0$ as
 \begin{align}
    h_{0} = \frac{p_{\rho}^{2}}{2m}+V_{\eff}(\rho) = E
\end{align}
where
\begin{align}
        V_{\eff} (\rho)= \frac{J^2}{2m\rho^{2}} - \frac{kMm}{\rho}
\end{align}
can be interpreted as an effective potential.
If the total energy of the system is close to the minimum of $V_{\eff} (\rho)$, it makes sense to think that a second order approximation of this potential (harmonic potential) describes reasonably well the gravitational interaction with the star.

Denote by $R = \frac{J^2}{k m^2 M}$  the value at which the minimum of the potential is attained.
A straightforward computation gives, introducing the dimensionless coordinate $\xi={\frac{\rho - R}{R}}$, that $V_{\eff} (\rho)$ can be rewritten in terms of $\xi$ as
\begin{align}\label{V}
    V(\xi) = \frac{1}{2} \frac{k M m}{R}\parens*{-1+ \frac{\xi^2}{(1+\xi)^2}}\ .
\end{align}
We will call the expansion of this potential in which we neglect the unessential constant $- \frac{1}{2} \frac{k M m}{R}$ and we keep only the second order term:
\begin{align}
    V_2(\xi) = \frac{1}{2} \frac{k M m}{R}\ \xi^2
\end{align}
the Gaussian approximation of the central interaction.
\begin{remark}
The Gaussian approximation is apparently a strong assumption, so we need a pair of comments. On one side, neglecting the first term in (\ref{V}) reflects the precise choice of  regarding the $R$'s as fixed quantities, rather than as thermodynamical variables (see also the next section).  
Secondly, for what concerns the approximation of  $V$ with its quadratic expansion, it will be clear (see Section \ref{sec: Planet} for a discussion) that in the applications of our model  to system of very small bodies (asteroids) such assumption is reasonable, because we will show that the interaction between the asteroids keeps the variance of $\xi$ of the same order of the unperturbed system, and the main terms in the corrections are related to configurations with small $\xi$, namely colliding asteroids. 
\end{remark}

\subsection{Free probability measure}
Denote by $R_{i}$ the radius of the $i$-th circular orbit, and consider the Gaussian approximation of its central potential.
\begin{align}
    V_{2,i}(\xi_i) -=\frac{1}{2} \frac{k M m_i}{R_i}\ \xi_i^2
\end{align}
To avoid heavy notations, we will denote such potential with $V_i(\xi_i)$, dropping the subscript $2$ until further notice.
We want to define a reference probability measure on the position of the body in the plane in absence of perturbations.
Recalling that $\xi_i$ and $\theta_i$ are, respectively, the dimensionless deviation of the distance
from the mean radius and the the true anomaly of the $i$-th body, we consider the probability measure
\begin{align}\label{eq:single_planet_measure_raw}
        \dmu_0(\xi_i, \theta_i) = \frac{e^{\beta_i V_i(\xi_i)}\de{\theta_i}\de{\xi_i}}{\int_{0}^{2\pi} \de{\theta_i} \int_{-\infty}^{\infty}\de{\xi_i}\ e^{-\beta_i V_i(\xi_i)}}.
\end{align}
where $\beta_i$ is a positive parameter.
Note that the kinetic part in the Hamiltonian doesn't play any role in the probability measure since it appears,
as a factor, both at the numerator and at the denominator. A similar fate would hold for the terms $-\frac{1}{2}\frac{kMm_i}{R_i}$ coming from (\ref{V}).

In statistical mechanics the parameter $\beta$ plays the role of the inverse temperature.
When the inverse temperature is large, the system tends to remain close to
the local minimizers of the Hamiltonian. Here each $\beta_{i}$ is determined so to have a probability distribution on the
unperturbed system having a standard deviation of the distance between the asteroid and the star much smaller than its average value. 
Since we assume small deviations from the average radius of the orbit the
harmonic approximation of the effective gravitational potential is reasonable.
The temperature $\beta_i$ has the dimension of the inverse of an energy. As already remarked, $\beta_i$ has to be rescaled by the inverse of the mass $m_i$ in order to be physically meaningful, see previous section. Moreover, the fact that our coordinates $\xi_i$ are dimensionless suggests to rescale the temperature by a factor $R_i$. Hence we choose
\begin{align}\label{eq:definition_of_beta}
        \beta_i = \frac{R_i}{k m_i M} \gamma_i^2
\end{align}
where $\gamma_i$ is a sufficiently large pure number, in order to have a small variance $\sigma^2(\xi_i)$ of the deviation $\xi_i$ from the average radius. As outlined in the introduction, $\gamma_i$ takes into account both the eccentricity and the interaction with planets. 

Introducing \eqref{eq:definition_of_beta} in \eqref{eq:single_planet_measure_raw} yields:
\begin{align}\label{eq:single_planet_measure}
        \dmu_0(\xi_i, \theta_i) 
                = \frac{\gamma_i}{(2\pi)^\frac{3}{2}}
                        e^{-\frac{1}{2} \gamma_i^2 \xi_i^2}\de{\theta_i}\de{\xi_i}
\end{align}
meaning that the measure of $\xi_i$ (without perturbations) is Gaussian with zero mean and variance
$\sigma_i^2 = \frac{1}{\gamma_i^2}$.

Note that, as a consequence of the previous considerations, in this model, each body has its own ``temperature'' $\beta_i$ that tunes its interaction with the star.

\subsection{Interacting probability measure}
When the interaction between the asteroids is taken into account, the probability distribution of the system is proportional to $e^{-H^{\mathrm{ad}}}$ where $H^{\mathrm{ad}}$ is the dimensionless Hamiltonian
\begin{align}\label{eq:Hamiltonian_adim_raw}
    H^{\mathrm{ad}}\parens*{\vec{\xi}, \vec{\theta}} =
    - \sum_{i=1}^{N} \frac{1}{2} \gamma_i^2 \xi_i^2 - \sum_{1\le i<j\le N} \beta_{ij} V_{ij}
\end{align}
with
\begin{align}
	V_{ij} = \frac{k m_i m_j}{\abs{\vec{x}_i - \vec{x}_j}}; \quad \vec{x}_i = \parens*{R_i(1 + \xi_i)\cos \theta_i; R_i(1 + \xi_i)\sin \theta_i}
\end{align}
and each $\beta_{ij}$ is a parameter tuning the interaction between the $i$-th and the $j$-th body.
Here the rescaling of $\beta_{ij}$ is not obvious, as in the case of the free measure, because $V_{ij}$ appears in the dynamics of both bodies. In order to have a measure related to the actual interaction among asteroids
it seems very reasonable that the dimensionless expression appearing eventually in our measure should respect
the following two conditions

1) The strength of the interaction has to be of the order of $m/M$, where $m$ is some kind of average between the masses of the bodies $i$ and $j$, as suggested by the gravitational nature of the interactions. 

2) The expression of the potential in terms of dimensionless units $\xi_i$ and $\xi_j$ shold be rescaled by a factor proportional to $R$, where again $R$ is some kind of average between the radius of the bodies $i$ and $j$.

To fulfill both request, we argue as follows. Other choices, fulfilling 1) and 2), would affect only the constants appearing in the subsequent estimates.

First of all, order the indices of the asteroids according to their average distance from the star, i.e. say that $R_i\le R_j$ if $i<j$. Consider the asteroids with indices $i$ and $j$ (with $i < j$) and consider the case where $\xi_i = \xi_j = 0$. In other words it means that the two planets have both distance from the star equal to the radii $R_i$ and $R_j$ respectively. Consider then the scenario where $\xi_i = \frac{1}{\gamma_i},\ \xi_j = \frac{1}{\gamma_j}$: each of the two asteroids has been moved away from the star by an amount equal to a free standard deviation. Call $\Delta V_i$ and $\Delta V_j$ the variation of the gravitational potential describing the interaction of the two bodies with the star associated with this change of scenario, and let $\Delta V_{ij}$ the corresponding change in the potential describing the gravitational interaction among the two planets.

In order to have a probabilistic weight due to the interaction among asteroids that is comparable with the one due to interaction with the star, we want that, when considering this change of scenario, the ratio $\frac{\Delta V_{ij}}{\Delta V_i + \Delta V_j}$ is the same as the ratio of the the corresponding variation in the exponent of $e^{-H^{\mathrm{ad}}}$, that is we want that
\begin{align}
    \frac{\Delta V_{ij}}{\Delta V_i + \Delta V_j}
        = \frac{\beta_{ij} \Delta V_{ij} }
               {\frac{1}{2}\gamma_i^2 \parens*{\frac{1}{\gamma_i}}^2 + \frac{1}{2}\gamma_j^2 \parens*{\frac{1}{\gamma_j}}^2 }
        =  \beta_{ij} \Delta V_{ij}.
\end{align}
Hence
\begin{align}
\beta_{ij}= \frac{1}{\Delta V_i + \Delta V_j}=\frac{1}{kM}\frac{R_i(1+\gamma_i)R_j(1+\gamma_j)}{m_iR_j(1+\gamma_j)+m_jR_i(1+\gamma_i)}
\end{align}

This means that
\begin{align}\label{Vij}
    \beta_{ij} V_{ij} = \gamma_{ij} \frac{\sqrt{R_i R_j}}{\abs{\vec{x}_i - \vec{x}_j}}:= \gamma_{ij} \frac{r_{ij}}{\abs{\vec{x}_i - \vec{x}_j}}
\end{align}
with
\begin{align}
    \gamma_{ij} = \frac{m_i m_j}{M}\frac{\sqrt{R_iR_j}(1+\gamma_i)(1+\gamma_j)}{m_iR_j(1+\gamma_j)+m_jR_i(1+\gamma_i)}\end{align}

The Statistical Mechanics model that we investigate is, therefore, defined through the following (dimensionless) Hamiltonian (calling again the dimensionless Hamiltonian and the dimensionless potential $H$ and $V$ respectively with an abuse of notation):

\begin{align}\label{eq:Hamiltonian_adim}
    H(\vec{\xi}, \vec{\theta}) =
    \sum_{i=1}^{N} \frac{1}{2} \gamma_{i}^2 \xi_i^2 - \sum_{1\le i<j\le N} \gamma_{ij} \frac{r_{ij}}{\abs{\vec{x}_i - \vec{x}_j}}=\sum_{i=1}^{N} \frac{1}{2} \gamma_{i}^2 \xi_i^2 - \sum_{1\le i<j\le N} V_{ij}
    \end{align}

Recall that the $\vec{x}_i$ are constrained by the hard core compatibility condition $\abs{\vec{x}_i - \vec{x}_j} \ge a_i + a_j$ where $a_i$ is the radius of the $i$-th body. Further note that, assuming the asteroids to have constant density $\delta$, we have $m_i = \frac{4}{3}\pi\delta a_i^3$.

The probability measure induced by the Hamiltonian \eqref{eq:Hamiltonian_adim} that we want to take into account to describe the planetary system is, therefore,
\begin{align}\label{eq:canonical_measure}
    \mu(\argph) = \frac{\int \de{\vec{\xi}} \int \de{\vec{\theta}} \, (\argph) \, e^{-H(\vec{\xi}, \vec{\theta})} }
                 {\int \de{\vec{\xi}} \int \de{\vec{\theta}} \, e^{-H(\vec{\xi}, \vec{\theta})} }.
\end{align}

Interpreting the variance of each $\xi_{i}$ as a quantity linked to the eccentricity of the $i$-th orbit, assessing the
stability of the system amounts to control the variance of the $\xi_{i}$'s.

In particular
we want to determine the conditions on $N$ and on the physical parameters (mass, radius of the orbits)
for which
the system is stable in the sense of the following
\begin{definition}
The system (\ref{eq:original_hamiltonian}) is called $\epsi -N$-stable if, for a fixed $\epsi$ and for all $i = 1, \ldots, N$
	\begin{align}
		\angbra{\xi_i^2} \le (1 + \epsi)\angbra{\xi_i^2}_{0}
	\end{align}
	where $\angbra{\xi_i^2}_{0}$ is the variance of $\xi_{i}$ with respect to $\dmu_{0}$.
\end{definition}

Indeed, if the previous condition is satisfied for an $\epsi$ small enough,
the deviations of the radii of the orbits of the asteroids,
with respect to the orbits they would have if the other asteroids were not there, stays small.

As outlined in the introduction, here there are a couple of delicate points deserving a discussion. First, what is the relation among this definition of stability and the evaluation of the stability of the orbit in an astronomical sense? Assume that an asteroid has a stable orbit for a time ${T}$ if its deviation from its average radius $R_i$ stays smaller than $A R_i$ for all $t<T$, with $A$ a suitable constant. Then we can deduce $T$ by the inverse of the probability $P(\xi_i>A)$ times the period of revolution $\tau$ of the asteroid around the star. In probability theory there are many ways to estimate $P(\xi_i>A)$. One of the rougher, needing just the control of the variance of $\xi$, is the Chebyshev inequality. A direct application of such inequality gives, for $A=1$, $T\approx \gamma_i^2 \tau$. This kind of estimates would hold for objects with masses comparable with the real masses of the asteroids in the main belt. However, this time (order of ten thousand years) is quite short in an astronomical sense. Our control of the variance, as it will turn out in the following sections, allows in principle to use different estimates for $P(\xi_i>A)$: for instance using the fact that the reference measure is Gaussian, one could use Chernoff inequality, or other methods involving the detailed control of higher order moments of the distribution. This could be easily done in principle, but it would strictly rely on the details of the reference measure. Recall that the standard statistical mechanics is based much more on the geometrical properties of the $N$-dimensional space, with $N$ of the order of the Avogadro's number, that on the details of the Gibbs probability measure. While the Chebyshev estimate mentioned above is quite robust, depending only on the fact that a reasonable reference probability should be strongly concentrated around its circular orbit, a more refined estimate will require some additional arguments regarding the faithfulness of the free reference measure. This will be the subject of further investigations.

The second point deserving a discussion is the role of the collisions in the evaluation of the canonical measure defined in
(\ref{eq:canonical_measure}). As it will be clear by the computations of the following subsection, the main problem of this approach is the control of the probability of the configuration in which $n$ asteroids are very close one with the other ($n$ body collisions), because the corresponding energy turns out to be negative and proportional to $n^2$. This gives rise to a probabilistic weight proportional to $C^{n^2}$, with $C>1$, and it is not clear how to control it from a combinatorial point of view. This seems to suggest that the collision terms are the leading one, and then only a very small amount of asteroids may be considered in order to have $\epsi -N$-stability with $\epsi$ reasonably small. A relatively standard but judicious control of the structure of the measure (\ref{eq:canonical_measure}), however, seems to indicate that the number of asteroids that can be taken into account in this model is not too different from the actual one (see section 4 below), and that this interpretation may suggest the fact that the number of asteroids that we see today is the relic of a much bigger initial set, in which a large part of asteroids has been lost due to the intrinsic instability of the system.

\subsection{Estimation of $\angbra{\xi_{m}^{2}}$}
The value of $\angbra{\xi_{m}^{2}}$ is given by
\begin{align}\label{eq:variance_xi_formal_expression}
    \angbra{\xi_{m}^{2}}
    & = \frac{\int \de{\vec{\xi}} \de{\vec{\theta}}\ \xi_m^2 e^{-H(\vec{\xi}, \vec{\theta})}  }
    		 {\int \de{\vec{\xi}} \de{\vec{\theta}} e^{-H(\vec{\xi}, \vec{\theta})} }
	  = \frac{\int \dmu_0(\vec{\xi}, \vec{\theta})\ \xi_m^2 e^{-V(\vec{\xi}, \vec{\theta})}}
	  		 {\int \dmu_0(\vec{\xi}, \vec{\theta}) e^{-V(\vec{\xi}, \vec{\theta})}}
\end{align}
where $H$ is defined in \eqref{eq:Hamiltonian_adim},  $\dmu_0(\vec{\xi}, \vec{\theta})$ is the product of the 
measures $\dmu_0\parens{\xi_i, \theta_i}$ defined in \eqref{eq:single_planet_measure} and
\begin{align}
	e^{-V(\vec{\xi}, \vec{\theta})} = e^{-\sum\limits_{\mathclap{i < j }} V_{ij}} =
	\smashoperator{\prod_{1\le i < j \le N}} e^{-V_{ij}} =
	\smashoperator{\prod_{1\le i < j \le N}} e^{\gamma_{ij}\frac{{r}_{ij}}{\abs{\vec{x}_i - \vec{x}_j}}}.
\end{align}
Note that the integral with respect to $\dmu_0(\vec{\xi}, \vec{\theta})$ must be restricted to ``compatible configurations'', that is, those configuration satisfying the hard core compatibility condition.

By writing
\begin{align}
	{\prod_{1\le i < j \le N}} e^{-V_{ij}} = {\prod_{1\le i < j \le N}} \left[\parens*{e^{-V_{ij}} - 1} + 1\right],
\end{align}
it is possible to rewrite \eqref{eq:variance_xi_formal_expression} exploiting
the following combinatorial identity
\begin{align}\label{eq:combinatorial_identity}
	\prod_{1 \le i < j \le N} (b_{ij} + 1) = \sum_{g \in \allgraphs{N}} \prod_{\edge{ij} \in E(g)} b_{ij},
\end{align}
where with $\allgraphs{N}$ we denote the set of all graphs with N vertices and with $E(g)$ the set of all edges of the graph $g$.
Thus we can write
\begin{align}\label{eq:varianceXiAllGraph}
	\angbra{\xi_m^2}
		= \frac{\int \dmu_0(\vec{\xi}, \vec{\theta})\ \xi_m^2 \prod\limits_{i<j} e^{-V_{ij}}}
	  		 {\int \dmu_0(\vec{\xi}, \vec{\theta}) \prod\limits_{i<j} e^{-V_{ij}}}
		= \frac{\sum\limits_{g \in \allgraphs{N}} \int \dmu_0(\vec{\xi}, \vec{\theta}) \, \xi_m^2 \prod\limits_{\edge{ij} \in E(g)} \parens*{e^{-V_{ij}} - 1}}
			   {\sum\limits_{g \in \allgraphs{N}} \int \dmu_0(\vec{\xi}, \vec{\theta}) \prod\limits_{\edge{ij} \in E(g)} \parens*{e^{-V_{ij}} - 1}}
\end{align}

Note that
\begin{align}
	\sum_{g \in \allgraphs{N}} \prod_{\edge{ij} \in E(g)} b_{ij}
	= \sum_{k=1}^{N} \sum_{X_1, \ldots, X_k} \prod_{l=1}^{k} \sum_{g \in \congraphs{X_l}} \prod_{\edge{ij} \in E(g)} b_{ij}
\end{align}
where $X_1, \ldots, X_k$ is a partition of the set $\braces{1, \ldots, N}$ and $\congraphs{X_l}$ is the set of all connected graphs with vertices in the set $X_l$.

Using this approach, and denoting by $X_{0}$ the component of the graph containing the vertex associated to the $m$--th body, \eqref{eq:varianceXiAllGraph} can be rewritten in terms of connected components in the following way

\begin{align} \angbra{\xi_m^2}\label{eq:varianceXiPeierlsArgument}
    & = \frac{{
    		\sum\limits_{k \ge 0} \, \sum\limits_{\substack{X_0, X_1, \ldots, X_k \\ \abs{X_l} \ge 2 \\  \abs{X_0} \ge 1}} \,
				\prod\limits_{l = 1}^{k}
    			\parens*{\sum\limits_{\mathclap{\phantom{xxx}g \in G_{X_l}}} \int \dmu_0(X_l)
               	                     \prod\limits_{\edge{ij} \in E(g)} \parens*{e^{-V_{ij}} - 1} }}{
            	\sum\limits_{\mathclap{\phantom{xx}g \in {G_{X_0 }}}} \int \dmu_0(X_0)
                	                 \, \xi_{m}^{2} \, \prod\limits_{\mathclap{\edge{ij} \in E(g)}} \parens*{e^{-V_{ij}} - 1}}
	          }{
	          \sum\limits_{k \ge 1} \sum\limits_{X_1, \ldots, X_k} \prod\limits_{l = 1}^{k}
    			\parens*{\sum\limits_{g \in G_{X_l}} \int \dmu_{0}(X_{l})
               	                     \prod\limits_{\edge{ij} \in E(g)} \parens*{e^{-V_{ij}} - 1} }
	          }
\end{align}

that allows to bound
$\angbra{\xi_m^2}$ as follows (this is what is commonly known as Peierls argument, see for instance \cite{peierls1936, gallavotti1972} for its definition in the case of low-temperature spin sytems)
\begin{align}\label{eq:varianceBoundPeierlsArgument}
	\angbra{\xi_m^2} & \le
            	\sum_{X \ni m}\sum\limits_{{g \in {G_{X}}}} \int \dmu_0(X)
                	                 \, \xi_{m}^{2} \, \prod\limits_{\mathclap{\edge{ij} \in E(g)}} \parens*{e^{-V_{ij}} - 1}
\end{align}
since
\begin{align}
 \frac{{
    		\sum\limits_{k \ge 0} \, \sum\limits_{\substack{X_0, X_1, \ldots, X_k \\ \abs{X_l} \ge 2 \\  \abs{X_0} \ge 1}} \,
				\prod\limits_{l = 1}^{k}
    			\parens*{\sum\limits_{\mathclap{\phantom{xxx}g \in G_{X_l}}} \int \dmu_0(X_l)
               	                     \prod\limits_{\edge{ij} \in E(g)} \parens*{e^{-V_{ij}} - 1} }}
	          }{
	          \sum\limits_{k \ge 1} \sum\limits_{X_1, \ldots, X_k} \prod\limits_{l = 1}^{k}
    			\parens*{\sum\limits_{g \in G_{X_l}} \int \dmu_{0}(X_{l})
               	                     \prod\limits_{\edge{ij} \in E(g)} \parens*{e^{-V_{ij}} - 1} }
	          }
 \le 1
\end{align}
Indeed, in the previous expression, since the pair potential $V_{ij}$ are negative in the integration region, the sums are over positive terms, both numerator and denominator are of the same type, but the denominator contains more terms.

We want to rewrite \eqref{eq:varianceBoundPeierlsArgument} in terms of a sum over trees instead of sum
over connected graphs using the so called Penrose Tree Graph identity introduced by Penrose in \cite{penrose67}, see also \cite{FernandezProcacci}. To this purpose, denoting $G_n$ the set of connected graphs on $n$ vertices and $T_n$  the set of trees, we first give the following
\begin{definition}
        A map $\mathfrak{M}: T_n \to G_n$ is called a partition scheme in  
        $G_n$ if, for all $\tau \in T_n$, $\tau \in \mathfrak{M}(\tau)$ and
        $G_n = \biguplus_{\tau \in T_n} \bracks{\tau, \mathfrak{M}(\tau)}$
\end{definition}
where $\biguplus$ denotes a disjoint union and
$\bracks{\tau, \mathfrak{M}(\tau)} = \braces{g \in G_n : \tau \subset g \subset \mathfrak{M}(\tau)}$
is a boolean interval with respect to the set-inclusion.
Further, given a partition scheme $\mathfrak{M}$ and a tree $\tau \in G_{n}$ write
$m(\tau) = E(\mathfrak{M}(\tau)) \setminus E(\tau)$ so that, in words, $m(\tau)$ represents the set of all edges that can be
added to $\tau$ to obtain a connected graph in the boolean interval $\bracks{\tau, \mathfrak{M}(\tau)}$.

With this notation, we have the following.
\begin{lemma}[General Penrose identity]\label{lem:penroseIdentity}
	Let $n > 2$ and let $\mathfrak{M}:T_{n} \to G_{n}$ be a partition scheme in $G_{n}$. Then
	\begin{align}
		\sum_{g \in G_{n}} \prod_{\edge{ij} \in E(g)} \parens*{e^{-V_{ij}} - 1}
		& = \sum_{\tau \in T_n} \prod_{\edge{ij} \in E(\tau)} \parens*{e^{-V_{ij}} - 1} \prod_{\edge{uv} \in m(\tau)} \parens*{e^{-V_{uv}}}
	\end{align}
\end{lemma}
whose proof is straightforward. Indeed:
\begin{proof}
Arguing as in \eqref{eq:combinatorial_identity}, we have
\begin{align}
	\sum_{g \in G_{n}} \prod_{\edge{ij} \in E(g)} \parens*{e^{-V_{ij}} - 1}
	& = \sum_{\tau \in T_n} \prod_{\edge{ij} \in E(\tau)} \parens*{e^{-V_{ij}} - 1} \sum_{S \subset m(\tau)} \prod_{\edge{uv} \in S} \parens*{e^{-V_{uv}} - 1} \\
	& = \sum_{\tau \in T_n} \prod_{\edge{ij} \in E(\tau)} \parens*{e^{-V_{ij}} - 1} \prod_{\edge{uv} \in m(\tau)} \parens*{e^{-V_{uv}} - 1} + 1 \\
	& = \sum_{\tau \in T_n} \prod_{\edge{ij} \in E(\tau)} \parens*{e^{-V_{ij}} - 1} \prod_{\edge{uv} \in m(\tau)} \parens*{e^{-V_{uv}}}
\end{align}
\end{proof}

Using again that $V_{ij} < 0$ in the integration region, we want to bound suitably the terms of the form 
$e^{-V_{ij}} - 1$. To do this we assume a priori that $ |V_{ij}|\le \frac{1}{2}$, we use the fact that $e^x-1\le \frac{5}{4}x$ for $0\le x\le \frac{1}{2}$, and then we verify a posteriori that the values of $ |V_{ij}|$ that guarantee the control of the $\epsi -N$-stability are smaller than $\frac{1}{2}$. We have
\begin{align}
	\angbra{\xi_{m}^{2}}
	& \le \sum_{n \ge 1} \sum_{\substack{\abs{X}=n \\ X \ni m}} \int \dmu_0(X)
		\xi_m^2 \,\sum_{g \in G_{n}} \prod_{\edge{ij} \in E(g)} \parens{e^{-V_{ij}} - 1} \\
	& = \sum_{n \ge 1} \sum_{\substack{\abs{X}=n \\ X \ni m}} \int \dmu_0(X) \xi_m^2
		\sum_{\tau \in T_n} \prod_{\edge{ij} \in E(\tau)} \parens{e^{-V_{ij}} - 1} \prod_{\edge{uv}\in m(\tau)} e^{-V_{uv}}\\
	& \le \sum_{n \ge 1} \sum_{\substack{\abs{X}=n \\ X \ni m}} \int \dmu_0(X) \xi_m^2
		\sum_{\tau \in T_n} \prod_{\edge{ij} \in E(\tau)} \frac{5}{4}\abs{V_{ij}} \prod_{\edge{uv} \in m(\tau)} e^{-V_{uv}} \label{eq:boundVarianceXi}
\end{align}

\section{Similar asteroids}
\label{sec:similar}

Here we consider the case of ``asteroids'' orbiting with similar radii and similar eccentricities around the star (the asteroids are in the same ``belt'').
Let $N$ be the total number of asteroids and let $R_{i} = R,\ \gamma_i=\gamma$ for all $i \in \braces{1, \ldots, N}$. Further let $a_{i}$ be the
the diameter of the $i$-th asteroid and let $a \le a_{i} \le 2a$.

We want to determine conditions ensuring the $\epsi -N$-stability of the system.
Assume $V_{ij} \le \frac{1}{2}$. It follows from \eqref{eq:boundVarianceXi}
\begin{align}\label{eq:ubound_orbit_radius_variance_similar_asteroids}
	\angbra{\xi_m^2} \le
	\angbra{\xi_m^2}_{0} \parens*{1 + \supnorm*{
		\sum_{n \ge 2} \sum_{\substack{\abs{X}=n\\ X \ni m}} \sum_{\tau \in T_{n}}
			\prod_{\edge{ij} \in \tau} \frac{5}{4} V_{ij}\prod_{\edge{ij}\in m(\tau)} e^{-V_{ij} } } }
\end{align}
where $\supnorm*{\cdot}$ is the supremum with respect to feasible configurations
and the addend $1$ represents the case $X_{0} = \{m\}$.
Denote by $\delta$ be the (common) density of the asteroids and by $\delta_{s}$ the density of the star.
The dimensionless potential can be written in this case in the form
\begin{align}\label{eq:pair_potential_reduced_mass}
        V_{ij} = - (\gamma+1)  \frac{m_i m_j}{M(m_i + m_j)} \frac{R}{\abs{\vec{x}_i - \vec{x}_j}}
\end{align}
and it is straightforward to verify that
\begin{align}\label{eq:ubound_pair_potential_similar_asteroids}
	\abs{V_{ij}} \le (\gamma+1) \frac{\delta}{\delta_s}  \frac{a_{<}^{3}}{a_{>}}\frac{R}{R_s^3}
\end{align}
with $a_{<} = \min\braces{a_{i}, a_{j}}$ and $a_{>} = \max\braces{a_{i}, a_{j}}$.

By \eqref{eq:ubound_orbit_radius_variance_similar_asteroids}, \eqref{eq:ubound_pair_potential_similar_asteroids} and the previous observations we have
\begin{align}
	\epsi \le \sum_{n \ge 2} \sum_{\substack{\abs{X}=n\\ X \ni m}} \sum_{\tau \in T_n}
		\parens*{(\gamma+1) \frac{\delta}{\delta_s} 5 a^{2}\frac{R}{R_s^3}}^{n-1}
		e^{{n^2 (\gamma+1)\frac{\delta}{\delta_s} 4 a^{2}\frac{R}{R_s^3}}}.
\end{align}\label{eq:bound_epsi_similar_asteroids}

Note that the fact that in this bound we have used \eqref{eq:ubound_pair_potential_similar_asteroids} means that we are considering the worst case scenario, i.e. an $n$-body collision, in order to evaluate the correction to the free measure due to the interaction among asteroids.

Introducing a dependence on $N$ and calling $A = N ((\gamma+1) \frac{\delta}{\delta_s}  5 a^{2}\frac{R}{R_s^3})$, $\bar A=\frac{4}{5}A$, \eqref{eq:bound_epsi_similar_asteroids} can be rewritten as
\begin{align}
	\epsi
		& \le \sum_{n\ge 2} \binom{N-1}{n-1} n^{n-2} \parens*{\frac{A}{N}}^{n-1} e^{\bar An} \\
        & = \sum_{n \ge 2} \frac{(N-1)(N-2)\cdots(N-n)}{(n-1)!} n^{n-2} \frac{A^{n-1}}{N^{n-1}}e^{\bar An}\\
        & \le  \sum_{n \ge 2} {A}^{n-1} e^{\bar An} \\
		& \le \sum_{n\ge 2} \parens*{Ae^{\bar A}}^{n-1} e^{\bar A} \\
        & \le e^{\bar A} \frac{Ae^{\bar A}}{1 - Ae^{\bar A}}
\end{align}
since, for $n\ge 2$
\begin{align}
	\frac{n^{n-2}}{\parens{n-1}^{n-1}}  \le 1,
\end{align}
\begin{align}
    k! \ge  \parens*{\frac{k}{e}}^k
\end{align}
and
\begin{align}
    \frac{(N-1)(N-2)\cdots(N-n)}{N^{n-1}} < 1.
\end{align}

Therefore, $\epsi -N$-stability of the system is guaranteed if $\frac{A e^{2\bar{A} }}{1 - Ae^{\bar A}} \le \epsi$. Rough numerical estimates show that if $A<1/5$ then $\epsi\le 2A$.

Note that in order to have $A=1/5$, $Na^2$ has to be bounded by a suitable constant. This means that, as outlined in the introduction, if $N$ increases the total mass of the asteroids, obviously proportional to $Na^3$, goes to zero.

\section{Asteroids with power-law mass distribution}
Now we consider a more realistic case (see for instance \cite{Bottke, Hughes, Ryan}): the $N$ asteroids have different masses/diameters (still under the assumption that they have common densities $\delta$) and may have different eccentricity. We will keep the assumption $R_i=R\ \forall i$ because the final estimate on $\epsi$ will represent an upper bound also for a different average radius $R_i$. Indeed, we are assuming that the deviations around the average radius have always a probabilistic weight of order 1, while a collision among asteroids, that gives the leading contribution in $V_{ij}$, has a free probability much smaller than 1 if the two average radii $R_i$ and $R_j$ are very different.

The distribution of the parameters $\gamma_i$ appearing in the free measure is also supposed to be not too spread: the eccentricity of the orbits may vary, but the perturbations due to the planets are similar for all the asteroids.

As far as the masses are concerned, we let the the diameters of the asteroids satisfy $a_{\min} \le a_{i} \le a_{\max}$  and we assume the following power-law  distribution for their diameters:
\begin{equation}
N(> a) = \frac{c}{a^{\nu}}
\end{equation}
where $N(> a)$ is the number of asteroids with diameter larger than $a$ and $c$ is a suitable constant. 
This law is assumed for the known asteroids belts in the solar system. To simplify our discussion, in the remainder we will set $\nu=2$. It will be clear that a different value of $\nu$ will affect only the constants, provided $\nu>1$.

We will define the unit of length in order to take $a_{\min} = 1$ and $a_{\max} = 2^{L}$ (for some natural number $L$). Note that $a_{\min} = 1$ implies $c = N$. In the applications describing the main asteroid belt 
in the Solar System the unit will be $1\,\mathrm{Km}$ (see below).

We partition the asteroids into $L$ classes $A_{1}, \ldots, A_{L}$. The $i$-th asteroid belongs to the $l$-th class if
$2^{l-1} \le a_{i} < 2^{l}$. In this case we write $i \in A_{l}$. Denoting by $N_{l}$ the number of asteroids in the $l$-th class
we have $N_{l} = N(> 2^{l-1}) - N(> 2^{l}) = \frac{3}{4^{l}}N$.

Let $i \in A_{l}$ and $j \in A_{m}$ with $l > m$. From \eqref{Vij}, setting 
$\gamma=\max_{1\le i\le N}\gamma_i$, it follows
\begin{align}
	\abs{V_{ij}} \le w_{lm}
		:= (\gamma+1) \frac{\delta}{\delta_{s}}  \frac{a_{m}^{3}}{a_{l}} \frac{R}{R_{s}^{3}}
		= (\gamma+1) \frac{\delta}{\delta_{s}}  4^{m} 2^{-(l-m-1)} \frac{R}{R_{s}^{3}}
\end{align}

For $l=m$ we have

\begin{align}
	w_{ll}= (\gamma+1) \frac{\delta}{\delta_{s}}  4^{l}  \frac{R}{R_{s}^{3}}
\end{align}
Hence
\begin{align}
    \epsi \le \sum_{n \ge 2} \sum_{\substack{n_1, \ldots n_l\\ \sum_i n_i = n}}
        \prod_{l=1}^{L} \binom{N_l}{n_l} n^{n-2}
        \max_{\tau} \prod_{\edge{ij} \in \tau} \parens*{e^{w_{lm}} - 1}
        \prod_{\edge{ij} \in m(\tau)} e^{w_{lm}}.
\end{align}

Since the estimates of the interactions $w_{lm}$ decay exponentially in $|l-m|$, the worst case is the tree $\tau$ having $n_L - 1$ connections among asteroids
in class $L$ and $n_l$ connections among asteroids of class $l$. Then

\begin{align}\label{eq:bound_worst_tree_mass_distributed_asteroids}
    \max_{\tau} \prod_{\edge{ij} \in \tau} \parens*{e^{w_{lm}} - 1}
        & \le  \parens*{(\gamma+1) \frac{\delta}{\delta_s} \frac{R}{R_s^3}4^L}^{n_L - 1}\prod_{l=1}^{L-1}\parens*{(\gamma+1) \frac{\delta}{\delta_s} \frac{R}{R_s^3}4^l}^{n_l}\\
        & \le A^{n-1}\parens*{\prod_{l=1}^{L-1} \parens*{\frac{1}{N_l}}^{n_l}}      \parens*{\frac{1}{N_L}}^{n_L - 1}
\end{align}

where in the last step we have set $A = \gamma \frac{\delta}{\delta_s} \frac{3R}{R_s^3}N$.

In addition, we have
\begin{align}
    \exp\braces*{\sum_{\phantom{xxxx}\mathclap{\edge{ij} \in m(\tau)}} w_{lm}}
        \le \exp\braces*{\sum_{l > m} n_l n_m w_{lm}}
        \le \exp\braces*{\sum_{l>m}\frac{A}{N_m}2^{-(l-m)}}.
\end{align}

Finally we obtain (assuming the asteroid $m$ is in class $1$):
\begin{align}
    \epsi & \le \sum_{n \ge 2}
                      \sum_{\substack{n_1 \ldots n_l\\ \sum_i n_i = n\\ n_i \ge 1}}
                      \prod_{l=2}^{L} \binom{N_l}{n_l} \binom{N_1 - 1}{n_1 - 1}
                        n^{n-2} A^{n-1} \prod_{l=1}^{L-1}
                            \parens*{\frac{1}{N_l}}^{n_l}\parens*{\frac{1}{N_L}}^{n_L -1} e^{nA}\\
                & \le e^{A+1}\sum{(ALe^{A})}^n
                    = e^{A+1}\frac{ALe^{A}}{1 - ALe^{A}}
\end{align}

\begin{remark}
It is interesting, in this slightly more realistic framework, to compare this result with the actual main belt of asteroids of Solar System.
The parameter $A$, setting $\gamma=50$, $\frac{\delta}{\delta_s}=2$ and the real values for $R$, $R_s$, has a value
$A\approx \frac{N}{5\times 10^5}$. Setting $L=10$, and considering only the asteroids with a diameter $a\ge 1\, \mathrm{Km}$, one finds that
to obtain $\varepsilon\le 1$ the condition on $A$ is $A\le 1/4$. This means that with our (rough) approximations $N\approx10^5$. The actual number of asteroids in the main belt having diameter larger that $1\, \mathrm{Km}$ is $N=10^6$
\end{remark}

\begin{remark}
The computation above assumes a minimal size of the asteroids. Here we present an indication of the fact that the power law mass distribution for the very light asteroids has to have an exponent $\nu<1$.
Calling $dN(a)$ the number of asteroids having the diameter between $a$ and $a+da$, we clearly have that,
if $N(> a) = \frac{N_1}{a^{\nu}}$, then
\begin{equation}\label{dienne}
dN(a)=N_1\nu \frac{da}{a^{\nu+1}}
\end{equation}
Considering that in the estimate of $\varepsilon$ we have to give a bound of the quantity $\sum_{ij}|V_{ij}|$
and using (\ref{eq:ubound_pair_potential_similar_asteroids}) and (\ref{dienne}) we get
\begin{equation}\label{sumv1}
\sum_{ij}|V_{ij}|\le (\gamma+1) \frac{\delta}{\delta_s} \frac{R}{R_s^3}\int_{a_{\rm min}}^{a_{\rm max}}da
\int_{a}^{a_{\rm max}}db N_1^2 \nu^2\frac{1}{a^{\nu+1}}\frac{1}{b^{\nu+1}}\frac{a^3}{b}
\end{equation}
It is now clear that $a_{\rm max}$ has to be such that $N(>a_{\rm max})=1$, and hence $a_{\rm max}=N_1^{\frac{1}{\nu}}$. Performing the elementary integrals in (\ref{sumv1}) we get
\begin{equation}\label{sumv2}
\sum_{ij}|V_{ij}|\le (\gamma+1) \frac{\delta}{\delta_s} \frac{R}{R_s^3}\frac{\nu^2N_1^2}{\nu+1}
\left[ \frac{N_1^{\frac{2-2\nu}{\nu}}-a_{\rm min}^{2-2\nu}}{ 2-2\nu}- \frac{N_1^{\frac{2-2\nu}{\nu}}-a_{\rm min}^{3-\nu}N_1^{-\frac{\nu+1}{\nu}}}{3-\nu}\right]
\end{equation}
This expression shows that if $a_{\rm min}$ is a finite value, say $a_{\rm min}=1$, then the conditions ensuring the control of $\varepsilon$ are $\nu>1$ and
\begin{equation}
(\gamma+1) \frac{\delta}{\delta_s} \frac{R}{R_s^3}\frac{\nu^2N_1^2}{\nu+1}<K
\end{equation}
with $K$ suitably chosen.
If we want to consider small $a_{\rm min}$, we have to assume that the distribution $N(> a) = \frac{N_1}{a^{\nu}}$, with $\nu>1$, is valid for $a>1$, while defining $N_<(>a)$ as the number of asteroids having a diameter between $a$ and 1, it has to be of the form
$N_<(> a) = \frac{N_1}{a^{\nu'}}$ with $\nu'<1$.
To our knowledge we do not have many observations on the mass distribution of the very small asteroids. However experimental data seems to show (see for instance \cite{Ryan}) that the exponent in the distribution tends to decrease for smaller asteroids.
\end{remark}

\section{Planets}\label{sec: Planet}
The basic idea developed in the previous sections is to describe the effect of the perturbation given by other distant objects, say planets, to the orbits of a large number of asteroids living in a single belt, i.e. with similar radii, by a probability distribution centered around a circular orbit. In this section we try to apply the same idea to a system of relatively few planets having well separated orbits.
In this case the free measure, i.e. the system obtained neglecting the interactions with the other planets, can be completely determined in terms of an elementary two-body problem.
However we shall see that a toy model in which $N$, the number of planets, is small ($\sim 10$), the masses of the planets may be quite different and the eccentricity of the orbits is very small (large $\gamma_i$) for all planets, keeps some interesting forecast performance, even when we substitute the well-known Keplerian orbit with a probability distribution.

The computations involved in this case, however, are quite different. Indeed, in the asteroids case the quantity to be controlled is the probability of collisions, and such collisions do not imply large deviations, in terms of the free measure, from the average value $R$ of the distance from the star, that is the same for all asteroids. In other words, the detailed structure of the free measure does not play any role, and the Gaussian approximation of the free measure is simply a way to compute very easily the free variance of the distribution of the distance from the star. The estimates, therefore, can be done always in the sense of an $L_\infty$ norm, and the fact that with a reasonable choice of the parameters we can keep $\varepsilon$ small means that the collisions give a negligible contribution to the interacting probability.

In the case of planets we will show that the system is $\epsi -N$-stable if the radii $R_i$ of the planets are very different, namely if the condition $R_{i} - R_{j} = c \parens{a^{i} - a^{j}}$ holds. This assumption amounts
to saying that the radii of the orbits satisfy the Titius--Bode law, that is, $R_{i} = b + c a^{i}$. Note that the Titius-Bode law is fulfilled quite accurately in the Solar System.

Since our main task is to verify that even in this case, with larger masses, the collisions give a negligible contribution to the interacting measure, we have to modify the previous computations: collisions are events with a very small probability with respect to the free measure, and hence we can not use $L_\infty$ estimates in order to evaluate the collisions. On the other side, since a collision is possible only when at least one planet has a very large fluctuation around his free orbit, the Gaussian approximation loses its meaning, and we need some initial estimates about the free complete measures.

The first important observation is that the probability density
\begin{equation}
dw_0(\xi,\theta)=\exp\left(-\frac{\gamma^2}{2}\frac{\xi^2}{(1+\xi)^2} \right)d\theta d\xi
\end{equation}
can not be normalized on the whole space. Indeed
\begin{equation}
\int_0^{2\pi}d\theta\int_{-1}^\infty d\xi \exp\left(-\frac{\gamma^2}{2}\frac{\xi^2}{(1+\xi)^2} \right)=\infty
\end{equation}

The simplest way out is to define the free measure on a finite space, say on a sphere of radius $2R_N$. This means that for the $i$-th planet $-1<\xi_i\le A_i=\frac{2R_N}{R_i}$. Since our task is to show that the collisions among planets have a negligible probability in the interacting measure, we will show that for large $\gamma$ the main contribution to the interacting measure will be given by the configurations in which each planet $i$ will have a distance from the star quite close to $R_i$, i.e. a $\xi_i$ of the order of $1/\gamma$. In other words, we are saying that a planet is inside the planetary system if it is not too far from the star.
Note that in the Solar System $A_{\rm Mercury}=200$.
Hence we will call
\begin{equation}\label{zeta}
Z_i=\int_0^{2\pi}d\theta_i\int_{-1}^{A_i}d\xi_i \exp\left(-\frac{\gamma_i^2}{2}\frac{\xi_i^2}{(1+\xi_i)^2} \right)
\end{equation}
Note that the main contribution in the integral comes from the interval $-1/2\le\xi_i\le1/2$ since the obvious $L_\infty$ estimate
\begin{equation}\label{zeta}
\int_{|\xi|>1/2}d\xi_i \exp\left(-\frac{\gamma_i^2}{2}\frac{\xi_i^2}{(1+\xi_i)^2} \right)\le A e^{-\frac{\gamma_i^2}{18}}
\end{equation}
holds.
It is a standard algebraic task, then, to show that for large values of $\gamma$ the variance of $\xi_i$ of the free measure for all planets $i$ is proportional to $\sigma_-^2=\frac{1}{4\gamma_i}$, as in the Gaussian approximation. Note that
\begin{equation}
\angbra{\xi_{i}^{2}}_{0}:=\int\dmu_0(\xi_i) \xi^{2} :=\frac{1}{Z_i}
\int_0^{2\pi}d\theta_i\int_{-1}^{A_i}d\xi_i\ \xi_i^2\exp\left(-\frac{\gamma_i^2}{2}\frac{\xi_i^2}{(1+\xi_i)^2} \right)
\end{equation}
and hence
\begin{equation}
\angbra{\xi_{i}^{2}}_{0}=\frac{1}{Z_i}
\int_0^{2\pi}d\theta_i\int_{|\xi_i|\le1/2}d\xi_i\ \xi_i^2\exp\left(-\frac{\gamma_i^2}{2}\frac{\xi_i^2}{(1+\xi_i)^2} \right) +
O(A_i^3e^{-\frac{\gamma_i^2}{18}})
\end{equation}
The interesting values for application to the Solar system are $A_i\le 200$ and $\gamma_i\ge 50$, and therefore the correction of order $A^3e^{-\frac{\gamma_i^2}{18}}$ is completely negligible.
Hence the leading part of the integral is dominated from above and from below by two Gaussian measures with variance
$\sigma_-^2=\frac{1}{4\gamma_i}$ and $\sigma_+^2=\frac{9}{4\gamma_i}$, respectively, and these are two bounds, both proportional to $\gamma_i^{-1}$, for the variance.

In what follows we will assume for simplicity that $\gamma_i=\gamma$ for all planets. For all $m=1,...,N$
we want to give an estimate of the quantity
\begin{align}\label{epspla}
	\epsi \angbra{\xi_{m}^{2}}_{0} =
		\sum_{n = 2}^{N} \sum_{\substack{\abs{X}=n\\ X \ni x_{m}}} \int \dmu_0(X) \xi^{2} \sum_{\tau \in T_{n}}
			\prod_{\edge{ij} \in \tau} \parens*{e^{-V_{ij}}-1}\prod_{\edge{ij}\in m(\tau)} e^{-V_{ij}}
\end{align}

where $V_{ij} = -2\gamma \frac{m_im_j}{M} \frac {R_iR_j}{R_im_j+R_jm_i}\frac{1}{\abs{\vec{x_i} - \vec{x_j}}}$

We call $\xi$ typical when $|{\xi_i}| < k\frac{1}{\gamma}$.
For a fixed $X$, we write $X = T \bigcup T^{c}$ with $T = \braces{i \in X | \xi_i \text{ is typical}}$.

Let us consider first the case in which $i$ and $j$ are both typical. Standard algebra shows that for $j>i$
\begin{align}
	\abs{\vec{x_i} - \vec{x_j}}
		& \ge R_{j}\parens*{1 - \frac{k}{\gamma} } - R_{i}\parens*{1 + \frac{k}{\gamma} } \\
				& \ge c_1 \parens*{a^j - a^i}
\end{align}
with $c_1 = c - \frac{2k}{\gamma} \parens{c + b}$

On the other side, if $i$ and/or $j$ are not typical
\begin{equation}
\abs{\vec{x_i} - \vec{x_j}}\ge r_j+r_i
\end{equation}
where $r_i$ is the radius of planet $i$.
To obtain an upper bound of $V_{ij} $ we observe that, recalling $j>i$
\begin{align}
	\frac {R_iR_j}{R_im_j+R_jm_i}\le\frac{b+ca^i}{m_{ij}^{\rm min}}
\end{align}
where obviously $m_{ij}^{\rm min}$ is the smallest mass between the planet $i$ and $j$.
If $i_{\rm min}$ is the smallest $i$ in the planetary system (recall for instance that in the Solar System Mercury corresponds to $i=-1$), calling $c_2=c+\frac{b}{a^{i_{\rm min}}}$ we obtain

\begin{align}
	\frac {R_iR_j}{R_im_j+R_jm_i}\le\frac{c_2a^i}{m_{ij}^{\rm min}}
\end{align}
Then if $i$ and $j$ are both typical, $j>i$, we get

\begin{equation}\label{vubar}
|V_{ij}|\le2\gamma\frac{{m_{ij}^{\rm max}}}{M}\frac{c_2}{c_1}\frac{1}{a^{j-i}-1}
\le\frac{2a\gamma}{a-1}\frac{m_{ij}^{\rm max}}{M}\frac{c_2}{c_1}a^{-(j-i)}:={\bar V}_{ij}
\end{equation}
Otherwise

\begin{equation}
|V_{ij}|\le{\bar V}_{ij}\frac{c_1(a^j-a^i)}{r_j+r_i}= {\widetilde V}_{ij}
\end{equation}

The strategy will be the following: first we evaluate in (\ref{epspla}) the case $X=T$, using for $V_{ij}$ the estimate ${\bar V}_{ij}$. We will call $\bar\epsi$ the estimate obtained in this way. In this case we will proceed as in the previous cases, with an $L_\infty$ estimate.

\begin{align}\label{estepspla}
	{\bar\epsi} \le&
		\sum_{n = 2}^{N} \sum_{\substack{\abs{X}=n\\ X \ni x_{m}}} \sum_{\tau \in T_{n}}
			\prod_{\edge{ij} \in \tau} \parens*{e^{{\bar V}_{ij}}-1}\prod_{i<j}e^{{\bar V}_{ij}}\\
		\le&	\left[ \prod_{i\ne m}\left( 1+\sum_{j\ne i}\left(e^{{\bar V}_{ij}}-1\right)\right)-1\right]\prod_{i<j}e^{{\bar V}_{ij}}
\end{align}
\begin{remark}
Note that in (\ref{estepspla}) we gave a quite rough estimate of the combinatorics on trees.
In particular we used that for trees rooted in $m$, since every vertex but $m$ has a unique ``predecessor'', $\sum_{\tau_X}\prod_{ij\in \tau}e_{ij}\le \prod_{i\ne m}\sum_{j\ne i}e_{ij}$. The addend $1$ takes into account the sum on $X$. The last addend $-1$ takes into account the fact that the sum in $n$ starts from 2. Since $N$ is small and the orbits are well separated this estimate is reasonable.
\end{remark}

Calling now
$c_3=\frac{2a\gamma}{a-1}\frac{m_{ij}^{\rm max}}{M}\frac{c_2}{c_1}$
we have ${\bar V}_{ij}\le c_3a^{-(j-i)}$. Assume $c_3<1/2$. Since $\sqrt{e}<5/3$ we have that
$e^{{\bar V}_{ij}}-1<\frac{5}{3}{\bar V}_{ij}$ and hence

\begin{align}\label{est2epspla}
{\bar\epsi} \le	\left[ \prod_{i\ne m}\left( 1+\frac{5}{3}c_3\sum_{j\ne i}a^{-|j-i|}\right)-1\right]e^{c_3\sum_{i< j}a^{-(j-i)}}
\end{align}

Using now the elementary inequalities $\sum_{i< j}a^{-(j-i)}\le\frac{1}{a-1}$ and $1+x\le e^x$ we finally get
\begin{align}\label{est3epspla}
{\bar\epsi} \le	\left( e^{\frac{5}{3}c_3N\frac{2}{a-1}}-1\right)e^{N\frac{c_3}{a-1}}
\end{align}

This concludes the estimate for $T=X$. The crucial relation to control the general case is the following. Call $\widetilde E$ the set of pairs $i,j$ of planets such that for their estimate we can not use (\ref{vubar}) (collisions).
For a fixed $T^c$ the contribution to $\epsi$, that we will denote $\epsi(T^c)$ can be bounded by
\begin{align}\label{est3epspla}
\epsi(T_c)\le\int d\mu_0(T^c)\ \prod_{ij\in {\widetilde E}}e^{{\widetilde V}_{ij}}\parens*{e^{{\bar V}_{ij}}-1}^{-1}
\end{align}

To prove (\ref{est3epspla}) it is enough to observe that, for all $\tau$
\begin{align}
\prod_{\edge{ij} \in E(\tau)} \parens{e^{-V_{ij}}-1} \prod_{\edge{ij}\in m(\tau)} e^{-V_{ij} }\le
\prod_{\edge{ij} \in E(\tau)} \parens{e^{{\bar V}_{ij}}-1} \prod_{\edge{ij}\in m(\tau)} e^{{\bar V}_{ij} }\prod_{ij\in {\widetilde E}}\frac{e^{{\widetilde V}_{ij}}}{\parens*{e^{{\bar V}_{ij}}-1}}
\end{align}
and then bound with $1$ the contribution of the integral $\int d\mu_0(T)$

The idea is then to bound the very large contribution due to $\prod_{ij\in {\widetilde E}}e^{{\widetilde V}_{ij}}\parens*{e^{{\bar V}_{ij}}-1}^{-1}$ with the smallness of $\int d\mu_0(T^c)$.

Let us start with the simplest case in which $T^c=\{i\}$ and the collision is with planet $i+1$. In this case the only estimate we can do for the probability of collision with respect to the free measue is the probability of $i$ to be non typical, $\mu_0(T^c)\le e^{-\frac{2k^2}{9}}$. On the other side the weight in the interacting measure of the collision is proportional to
$e^{{\bar V}_{i,i+1}\frac{c_1a^i(a-1)}{r_i+r_{i+1}}}\approx e^{c_3\frac{c_1a^i(a-1)}{a(r_i+r_{i+1})}}$.
Hence our condition in order to control the single collisions will be
\begin{equation}\label {cond}
\frac{2k^2}{9}>c_3\frac{c_1a^i(a-1)}{a(r_i+r_{i+1})}>\frac{2a\gamma}{a-1}\frac{m_{ij}^{\rm max}}{M}c_2\frac{a^i}{r_i+r_{i+1}}
\end{equation}
We outline that for $N$ not too large, say $N\ge 10$, the case $T^c=\{i\}$ is the leading one: in order to evaluate the $l$-body collisions the contribution
$c_3\frac{c_1a^i(a-1)}{a(r_i+r_{i+1})}$ has to be multiplied by ${l\choose 2}$, while the contribution $\frac{2k^2}{9}$ becomes much larger, because at least $l-2$ planets have to undergo a deviation in $\xi$ of order $1$, and hence the factor becomes of the order of $\gamma^2$ instead of $k^2$.

We end this section outlining that (\ref{cond}) and (\ref{est3epspla}) can be specified in the case of the planets of the Solar System and in the case of the Galilean satellites. Note, however, that the numerical estimates we stated in the generic case may be specified better once we know the actual value of the parameter. In the case of the planets we solve simply both conditions, in the sense of the $1$-stability, using as free parameter $m^{\rm max}$. $k$ can be chosen in order to have the largest possible value of $m^{\rm max}$. Reasonable values for the parameters are:

\begin{itemize}
\item$\gamma=150$, since the eccentricity of the orbits are very small.
\item$k=30$,
\item$c_2=1 UA$
\item $a=2$
\item $a^{i^{\rm max}}= 128$
\end{itemize}

Then it is possible to satisfy  (\ref{cond}) and (\ref{est3epspla}) in order to have $\epsi<1$ with a value of $m^{\rm max}$ similar to the Earth's one.
In the case of Galilean satellites, in which $N=4$ and, most of all, $b=0$, we can control the various steps of the estimates much better. The combinatorics on trees and the sums on $V_{ij}$ can be written more explicitly, obtaining eventually that a ratio $m^{\rm max}/m_J\approx 10^{-4}$, which is the actual value, ensures $1$-stability.

\section{Conclusions and open problems}
The aim of this work is to outline the fact that with a judicious but quite standard use of results typical of equilibrium statistical mechanics one can evaluate some global features of the systems of particles rotating around a much bigger body. The estimates presented here are quite rough, and they can be surely improved by a careful numerical evaluation of the constants appearing in the theory. Nevertheless, the results we got, namely an evaluation of the ``thermodynamical'' stability of the main asteroid belt, of the planets in the solar system and of the Galilean satellites, give quantitative estimates not too distant from the real data, and seem therefore to indicate that this approach to the planetary system gives a reasonable possibility to understand the global structure of the Solar system. More precisely, our model seems to indicate that in order to have a thermodynamically stable system the masses of the particles orbiting around the fixed large body have to be very small if the orbit's parameters of the particle are similar, but they can increase if the objects are far apart. It would be nice to have some data about the very small objects in the belts of the Solar system (main belt of asteroids, trans--Neptunian belts, rings around the planets) because our model seem to indicate that the distribution of the very light objects in a belt has to have a different scaling law with respect to the one of the heavier ones. 

\section*{Acknowledgments}
We are grateful to Antonio Ponno and Zacharias Roupas for suggesting valuable references.
G.P. has been supported by the H2020 Project Stable and Chaotic Motions in the Planetary Problem (Grant 677793 StableChaoticPlanetM of the European Research Council).
B.S. acknowledges the MIUR Excellence Department Project awarded to the Department of Mathematics, University of Rome Tor Vergata, CUP E83C18000100006, the PRIN I-CELMECH funded by the MIUR, and the A*MIDEX project (n. ANR- 11-IDEX-0001-02) funded by the “Investissements d’Avenir” French Government program, managed by the French National Research Agency (ANR).
A.T. has been supported by the MIUR Project FARE 2016 (Grant R16TZYMEHN, Gravitational Systems Dynamics) until May 2020 and by the H2020 Project Stable and Chaotic Motions in the Planetary Problem (Grant 677793 StableChaoticPlanetM of the European Research Council) from October 2020.

\end{document}